\documentclass[aps,pre,reprint,groupedaddress]{revtex4-1}
\usepackage{amsmath}
\usepackage{amsfonts}
\usepackage{graphicx}
\begin{document} 
\title{Zero-range process with finite compartments: Gentile's statistics and glassiness}
\author{Artem Ryabov} 
\email[]{rjabov.a@gmail.com}
%\thanks{}
\affiliation{Charles University in Prague, Faculty of Mathematics and Physics, Department of Macromolecular Physics, V Hole{\v s}ovi{\v c}k{\' a}ch 2, 180~00 Prague, Czech Republic}
\date{\today} 
%%%%%%%%%%%%%%%%%%%%%%%%%%%%%%%%%%%%%%%%%%%%%%%%%%%%%%%%%%%%%%%%%%%%%%%%%%%%%%%%%%%%%%%%%%%%%%%%%%%%%%%%%%%%%%%%%%%%%%%%%%%%%%%%%%%%%%%%%%%
%%%%%%%%%%%%%%%%%%%%%%%%%%%%%%%%%%%%%%%%%%%%%%%%%%%%%%%%%%%%%%%%%%%%%%%%%%%%%%%%%%%%%%%%%%%%%%%%%%%%%%%%%%%%%%%%%%%%%%%%%%%%%%%%%%%%%%%%%%%
%%%%%%%%%%%%%%%%%%%%%%%%%%%%%%%%%%%%%%%%%%%%%%%%%%%%%%%%%%%%%%%%%%%%%%%%%%%%%%%%%%%%%%%%%%%%%%%%%%%%%%%%%%%%%%%%%%%%%%%%%%%%%%%%%%%%%%%%%%%
\begin{abstract} 
We discuss statics and dynamics of condensation in a zero-range process with compartments of limited sizes. For the symmetric dynamics the stationary state has a factorized form. For the asymmetric dynamics the steady state factorizes only for special hopping rules which allow for overjumps of fully occupied compartments. In the limit of large system size the grand canonical analysis is exact also in a condensed phase, and for a broader class of hopping rates as compared to the previously studied systems with infinite compartments. The dynamics of condensation exhibits dynamical self-blocking which significantly prolongs relaxation times. These general features are illustrated with a concrete example: an inhomogeneous system with hopping rates that result in Bose-Einstein-like condensations.
\end{abstract} 
% \pacs{}
%\keywords{} 
\maketitle 
%%%%%%%%%%%%%%%%%%%%%%%%%%%%%%%%%%%%%%%%%%%%%%%%%%%%%%%%%%%%%%%%%%%%%%%%%%%%%%%%%%%%%%%%%%%%%%%%%%%%%%%%%%%%%%%%%%%%%%%%%%%%%%%%%%%%%%%%%%%
%%%%%%%%%%%%%%%%%%%%%%%%%%%%%%%%%%%%%%%%%%%%%%%%%%%%%%%%%%%%%%%%%%%%%%%%%%%%%%%%%%%%%%%%%%%%%%%%%%%%%%%%%%%%%%%%%%%%%%%%%%%%%%%%%%%%%%%%%%%
%%%%%%%%%%%%%%%%%%%%%%%%%%%%%%%%%%%%%%%%%%%%%%%%%%%%%%%%%%%%%%%%%%%%%%%%%%%%%%%%%%%%%%%%%%%%%%%%%%%%%%%%%%%%%%%%%%%%%%%%%%%%%%%%%%%%%%%%%%%

%%%%%%%%%%%%%%%%%%%%%%%%%%%%%%%%%%%%%%%%%%%%%%%%%%%%%%%%%%%%%%%%%%%%%%%%%%%%%%%%%%%%%%%%%%%%%%%%%%%%%%%%%%%%%%%%%%%%%
%%%%%%%%%%%%%%%%%%%%%%%%%%%%%%%%%%%%%%%%%%%%%%%%%%%%%%%%%%%%%%%%%%%%%%%%%%%%%%%%%%%%%%%%%%%%%%%%%%%%%%%%%%%%%%%%%%%%%
%%%%%%%%%%%%%%%%%%%%%%%%%%%%%%%%%%%%%%%%%%%%%%%%%%%%%%%%%%%%%%%%%%%%%%%%%%%%%%%%%%%%%%%%%%%%%%%%%%%%%%%%%%%%%%%%%%%%%
\section{Introduction}
One of the most intriguing abilities of statistical physics is to explain phase transitions of macroscopic systems starting from microscopic local dynamical rules.  While for systems in thermal equilibrium the equilibrium statistical mechanics provides a satisfactory insight, for far-from-equilibrium systems we still lack a unifying theory. Consequently, the nonequilibrium systems must be studied one by one with only few rare examples known to be analytically tractable. 

One such exception is a zero-range process (ZRP), a stochastic interacting-particle model defined on a lattice. The basic feature of ZRP is that the hopping rates of particles depend only on the departure site. ZRP was introduced by Spitzer \cite{Spitzer} in 1970, however, recently it attracted considerable attention of the statistical physics community. The interest in the model stems from its numerous applications, e.g. to clustering in shaken granular gases, traffic flow, condensation on networks, in macroeconomies (for a review see \cite{EvansZRP2, EvansZRP1, GodrecheReview}). From the fundamental viewpoint the model allows for rigorous  study of different types of condensation transitions including, e.g., the effects of disorder (for a review see \cite{GodrecheDisorder}).

In the present paper we add a new element into the dynamics of ZRP: we assume that each lattice site (or compartment) \emph{has a finite capacity}, i.e., it can hold only a finite number of particles. The finite site capacity implies that a) in the limit of large system size the grand canonical analysis becomes exact for any density of particles (in contrast to the infinite-capacity ZRP where the grand-canonical ensemble often fails to describe the condensed phase \cite{CanonicalCondensation, CanonicalCondensationII}), b) we are able to treat analytically a broader class of hopping rates (than in the infinite-capacity case), c) the dynamics of condensate growth exhibits a dynamical self-blocking which significantly  prolongs relaxation times (the entropic effect known from models of glasses \cite{GlassBiroli, RitortReview, BackgamonRitort, SlaninaChvosta, Sherrington, GodrecheReviewDynamics, KobAndersen, Sellitto}). 

We proceed as follows. In Sec. \ref{SymmetricD} the steady state for the symmetric dynamics is discussed. Asymmetric jumping rules for which the steady state has a factorized form are defined in Sec. \ref{AsymmetricD}.  Equivalence of ensembles is proved in Sec. \ref{ensembles}. As a particular example, we consider statics and dynamics of ZRP with the hopping rates that lead to Bose-Einstein condensation in infinite-capacity ZRP \cite{EvansBE, AngelZRP} in Sec. \ref{example}.  

%%%%%%%%%%%%%%%%%%%%%%%%%%%%%%%%%%%%%%%%%%%%%%%%%%%%%%%%%%%%%%%%%%%%%%%%%%%%%%%%%%%%%%%%%%%%%%%%%%%%%%%%%%%%%%%%%%%%%
%%%%%%%%%%%%%%%%%%%%%%%%%%%%%%%%%%%%%%%%%%%%%%%%%%%%%%%%%%%%%%%%%%%%%%%%%%%%%%%%%%%%%%%%%%%%%%%%%%%%%%%%%%%%%%%%%%%%%
\section{\label{SymmetricD} Symmetric dynamics}

For simplicity we consider a one-dimensional lattice containing $L$ sites labelled $i=1,\ldots,L$, with periodic boundary conditions and we assume nearest-neighbor particle hopping. The number of particles at the site $i$, $n_{i}$, is integer and it is bounded by  
\begin{equation}
\label{capacity}
0\leq n_{i}\leq C_{i}, \quad i=1,\ldots,L,
\end{equation}
where the integer $C_{i}$ equals the capacity of the  $i$-th site. Particle hopping is \emph{symmetric}. The rates with which a single particle leaves the $i$-th site and arrives at the site $(i-1)$ ($W_{i}^{\rm L}(n_{i})$), or at the site  $(i+1)$ ($W_{i}^{\rm R}(n_{i})$) are given by 
\begin{eqnarray} 
\label{rate}
W_{i}^{\rm L}(n_{i}) &=& u_{i}(n_{i})\, \theta\!\left(C_{i-1}-n_{i-1}\right),\\
W_{i}^{\rm R}(n_{i}) &=& u_{i}(n_{i})\, \theta\!\left(C_{i+1}-n_{i+1}\right),
\end{eqnarray}
where the unit step theta functions prevent the particle from hopping on the fully occupied site. Probability $P_{t}(\mathbf{n})$ of finding the system in a configuration $\mathbf{n}=(n_{1},\ldots,n_{L})$ satisfies the master equation
\begin{eqnarray}  
\label{ME}
 && \frac{{\rm d}P_{t}(\mathbf{n})}{{\rm d} t} =
\sum_{i=1}^{L} \theta\left(n_{i}\right) \left\{ W_{i-1}^{\rm R}(n_{i-1}+1) P_{t}(\mathbf{n}_{i-1,i})+
\right. \\ 
\nonumber 
&& + \left. \! 
 W_{i+1}^{\rm L}(n_{i+1}\!+\!1) P_{t}(\mathbf{n}_{i+1,i}) 
\! -\! \left[ W_{i}^{\rm L}(n_{i})\! + W_{i}^{\rm R}(n_{i}) \right]\! P_{t}(\mathbf{n})  \right\}, 
\end{eqnarray} 
where the configuration $\mathbf{n}_{i-1,i}$ ($\mathbf{n}_{i+1,i}$) is identical to the configuration $\mathbf{n}$ except for the exchange of a single particle between the sites $(i-1)$ and $i$ ($(i+1)$ and $i$), namely, 
$\mathbf{n}_{i-1,i}=(\ldots,n_{i-1}+1,n_{i}-1,\ldots)$,
$\mathbf{n}_{i+1,i}=(\ldots,n_{i}-1, n_{i+1}+1,\ldots)$.
The equilibrium solution of the master equation (\ref{ME}) is found in the product form  
\begin{equation} 
\label{probab}
P(\mathbf{n})=\frac{1}{Z_{L,N}}\prod_{i=1}^{L} f_{i}(n_{i}),
\end{equation}
where the single-site statistical weights $f_{i}(n_{i})$ are expressed in terms of hopping rates: 
\begin{equation} 
\label{weights}
f_{i}(n_{i}) = \prod_{n=1}^{n_{i}}\frac{1}{u_{i}(n)},\quad f_{i}(0)=1.
\end{equation} 
The normalization $Z_{L,N}$ is computed by summing the product $\prod_{i=1}^{L} f_{i}(n_{i})$ over all configurations ${\bf n}$ compatible with two constraints: the first is expresses in (\ref{capacity}), the second is the conservation of the total number of particles $N$, $N\!=\!\sum_{i=1}^{L}n_{i}$.

It is straightforward to verify that,  \emph{for any site capacities} $C_{i}$, $i=1,\ldots,L$, the above equilibrium distribution cancels individually each summand on the right-hand side of the master equation (\ref{ME}). Symmetry of the dynamics \emph{is necessary} for this cancellation. Furthermore, provided that the dynamics is symmetric, the above equilibrium distribution solves the stationary master equation on an arbitrary lattice.

%%%%%%%%%%%%%%%%%%%%%%%%%%%%%%%%%%%%%%%%%%%%%%%%%%%%%%%%%%%%%%%%%%%%%%%%%%%%%%%%%%%%%%%%%%%%%%%%%%%%%%%%%%%%%%%%%%%%%
%%%%%%%%%%%%%%%%%%%%%%%%%%%%%%%%%%%%%%%%%%%%%%%%%%%%%%%%%%%%%%%%%%%%%%%%%%%%%%%%%%%%%%%%%%%%%%%%%%%%%%%%%%%%%%%%%%%%%
\section{\label{AsymmetricD} Asymmetric dynamics}

In contrast to the symmetric dynamics studied in Section \ref{SymmetricD}, ZRP with asymmetric dynamics (e.g. frequently studied totally asymmetric case) and with \emph{nearest-neighbor} particle hopping does not seem to possess a factorized steady state when finite site capacities are assumed. However, if we relax the assumption of the nearest-neighbor hopping, the factorized steady state can be recovered even for asymmetric dynamics. 

In the following we consider the totally asymmetric dynamics on a one-dimensional lattice of $L$ sites with periodic boundary conditions. The sites are labeled from left to right. The number of particles at the site $i$, $n_{i}$, is bounded by the maximum site capacity $C_{i}$ in accordance with (\ref{capacity}). A single particle departs from the site $i$  with the rate $u_{i}(n_{i})$ and it hops to the right. The arrival site is not necessarily the site number $(i+1)$, instead it is chosen according to the following \emph{jump-over policy}. The arrival site \emph{is the closest site} to the right of $i$ that is not fully occupied by particles. 

Let $\lambda(i)$ be the label of the closest site \emph{to the left} of $i$ that is not fully occupied by particles ($n_{\lambda(i)}<C_{\lambda(i)}$). Then the master equation for the probability that, at the time $t$, the system is in the configuration ${\bf n}$ reads 
\begin{equation}
\begin{split}
\label{MEasy} 
\frac{{\rm d} P_{t}({\bf n})}{{\rm d}t} = 
\sum_{i=1}^{L}
\theta(n_{i})\! & \left[ u_{\lambda(i)}(n_{\lambda(i)}\!+\!1) P_{t}({\bf n}_{\lambda(i),i})
-\right. \\
& \left. \,\, - u_{i}(n_{i})P_{t}({\bf n})  \right]. 
\end{split}
\end{equation}
On the right-hand side of Eq.\ (\ref{MEasy}), each summand accounts for one gain and for one loss term. The gain term is due to a possible particle jump from a \emph{uniquely chosen} site $\lambda(i)$. The jump changes the system configuration ${\bf n}_{\lambda(i),i}=(\ldots,n_{{\lambda(i)}}+1,\ldots,n_{i}-1,\ldots)$ to the configuration ${\bf n} =(\ldots,n_{{\lambda(i)}},\ldots,n_{i},\ldots)$. The loss term is due to the possible particle hopping from the site $i$ on a \emph{uniquely chosen} 
arrival site to the right of $i$ \footnote{The fact that the both sites are uniquely defined for any $i$ ensures that the detailed balance holds. In order to extend the present asymmetric model to a more general lattice we would have to define a proper jump-over policy on the corresponding graph.}. The master equation (\ref{MEasy}) has the above form for any system configuration ${\bf n}$ where $i\neq \lambda(i)$ for $i=1,\ldots,L$. If there exists $i$ such that $i=\lambda(i)$, then we assume that the $i$-th summand in (\ref{MEasy}) is identically equal to zero. This ensures that the arrival site is always different from the departure site (notice that for $i=\lambda(i)$ the only site that is not fully occupied is the $i$-th one). 

The factorized form (\ref{probab}) with statistical weights (\ref{weights}) cancels individually each summand on the right-hand side of the master equation  (\ref{MEasy}). Hence Eqs.\ (\ref{probab}), (\ref{weights}) gives us the steady state probability distribution also for the present totally asymmetric model.

%%%%%%%%%%%%%%%%%%%%%%%%%%%%%%%%%%%%%%%%%%%%%%%%%%%%%%%%%%%%%%%%%%%%%%%%%%%%%%%%%%%%%%%%%%%%%%%%%%%%%%%%%%%%%%%%%%%%%
%%%%%%%%%%%%%%%%%%%%%%%%%%%%%%%%%%%%%%%%%%%%%%%%%%%%%%%%%%%%%%%%%%%%%%%%%%%%%%%%%%%%%%%%%%%%%%%%%%%%%%%%%%%%%%%%%%%%%
\section{Grand canonical analysis}
\label{ensembles}
  
In order to derive any quantity of interest using the joint distribution (\ref{probab}) it is convenient to work within the grand canonical ensemble (see \cite{EvansZRP2, EvansZRP1, GodrecheReview}). That is, instead of the total number of particles $N$, $N\!=\!\sum_{i=1}^{L}n_{i}$, we fix a fugacity $z$. The fugacity determines the particle density $\rho = N/L$ through 
\begin{equation} 
\rho(z) = \frac{1}{L} \sum_{i=1}^{L} \left< n_{i} \right>,
\end{equation}
where the average is taken with respect to the grand canonical probability distribution: 
\begin{eqnarray}
\label{probabGC} 
P_{\rm GC}(\mathbf{n}) &=& \prod_{i=1}^{L}\frac{f_{i}(n_{i})z^{n_{i}}}{q_{i}(z)}, 
\quad q_{i}(z) = \sum_{n=0}^{C_{i}} f_{i}(n)z^{n}.
\end{eqnarray} 
The canonical and the grand canonical ensembles are equivalent in the thermodynamic limit: $L\to \infty, N\to \infty$, $\rho$ fixed. The equivalence is proved in Section $4$ of Ref. \cite{EvansZRP2}. Since, in the present case, the grand canonical partition function of the whole system, $Q(z) = \prod_{i=1}^{L}q_{i}(z)$, is a polynomial of a finite degree $\left(\sum_{i=1}^{L}C_{i}\right)$, the saddle point approximation \cite{EvansZRP2} is valid for any particle density $\rho$, $\rho \in (0,\sum_{i=1}^{L}C_{i}/L)$, and for arbitrary hopping rates $u_{i}(n)$. 

An important consequence emerges: if the hopping rates are such that the system can exist in different phases, then (in the limit of large system size, and provided that all $C_{i}$ are finite) the equivalence of ensembles holds \emph{in all phases}. Hence the finite site capacities regularize the grand canonical ensemble which otherwise frequently fails to describe the condensed phase in infinite-capacity ZRP. 

%%%%%%%%%%%%%%%%%%%%%%%%%%%%%%%%%%%%%%%%%%%%%%%%%%%%%%%%%%%%%%%%%%%%%%%%%%%%%%%%%%%%%%%%
\begin{figure}[t!]
\includegraphics[width=0.4\textwidth]{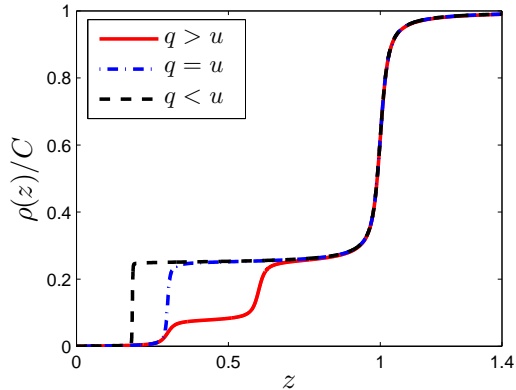}
\centering
\caption{\label{fig:densityfugacityI} (Color online) The scaled density of particles versus the fugacity $z$ as obtained from Eq.\ (\ref{densityfugacity}) for $u=0.3$, $C=200$, $M=60$, $L_{1}/L=1/4$ and for three different values of $q$: $q=u/2$ (the dashed line), $q=u$ (the dot-dashed line), $q=2u$ (the solid line).}
\end{figure} 
%%%%%%%%%%%%%%%%%%%%%%%%%%%%%%%%%%%%%%%%%%%%%%%%%%%%%%%%%%%%%%%%%%%%%%%%%%%%%%%%%%%%%%%% 

%%%%%%%%%%%%%%%%%%%%%%%%%%%%%%%%%%%%%%%%%%%%%%%%%%%%%%%%%%%%%%%%%%%%%%%%%%%%%%%%%%%%%%%%%%%%%%%%%%%%%%%%%%%%%%%%%%%%%
%%%%%%%%%%%%%%%%%%%%%%%%%%%%%%%%%%%%%%%%%%%%%%%%%%%%%%%%%%%%%%%%%%%%%%%%%%%%%%%%%%%%%%%%%%%%%%%%%%%%%%%%%%%%%%%%%%%%%
\section{\label{example} Example}
\subsection{Statics}

As the simplest nontrivial example let us now study the steady state of the inhomogeneous system corresponding to hopping rates 
\begin{eqnarray} 
\label{BErates}
u_{i}(n)&=&
\left\{
\begin{array}{ll}
u(n), & {\rm for} \,\, i \in [1,L_{1}],  \\[4pt]
1, & {\rm for} \,\, i \in [L_{1}+1,L]. 
\end{array}\right. 
\end{eqnarray} 
We assume that the capacities of all sites are equal to $C$. The particle-dependent rate $u(n)$ equals to $u$, for $n \in [1,M]$, 
and it equals to $q$, for $n \in [M+1,C]$. Further we always assume that $q<1$, $u<1$. In other words, the lattice (the ring) consists of two homogeneous domains: ``the slow domain'' (sites labeled by $i=1,\ldots,L_{1}$ with the particle hopping rate $u(n)$, $u(n)<1$) and ``the fast domain'' (sites $i=L_{1}+1,\ldots,L$, the hopping rate equals $1$). In all illustrations we take $L_{1}/L=1/4$.

Interestingly enough, in the infinite-capacity ZRP, the hopping rates (\ref{BErates}) lead to the phase transition analogous to Bose-Einstein condensation of an ideal Bose gas \cite{EvansZRP1, EvansZRP2, EvansBE, AngelZRP, BWaclaw1, Bogacz}. The formal equivalence with the grand canonical equilibrium quantum statistics is achieved by setting $z/u_{i}(n) = {\rm e}^{-\beta(\varepsilon(i)-\mu)}$. For a finite $C$, (and for the hopping rates (\ref{BErates})) this substitution maps the probabilities (\ref{probabGC}) onto the equilibrium grand canonical distribution for  particles obeying \emph{intermediate statistics} which was introduced by Gentile in 1940 \cite{Gentile40}. The intermediate statistics interpolates between the Fermi-Dirac ($C\!=\! 1$) and the Bose-Einstein ($C\! = \! \infty$) cases (see also  \cite{Cattani} for an overview, \cite{Lavenda, terHaar1952199} for criticism, \cite{Niven} for occurrence in urn models, and \cite{Tsallis, Avraham} for thermodynamic properties of ``paragas''). 

For the hopping rates (\ref{BErates}), the density of particles $\rho$ and the fugacity $z$ are related through
\begin{align} 
\nonumber
 \rho(z)= & 
\left(\frac{L_{1}}{L}\right) \frac{\sum_{n=0}^{M}n \left(\frac{z}{u}\right)^{n}+\left(\frac{q}{u}\right)^{M} 
\sum_{n=M+1}^{C}n \left(\frac{z}{q}\right)^{n}}
{\sum_{n=0}^{M}\left(\frac{z}{u}\right)^{n} + \left(\frac{q}{u}\right)^{M} \sum_{n=M+1}^{C} \left(\frac{z}{q}\right)^{n}}+\\
&  
\qquad + \left(1- \frac{L_{1}}{L} \right) \frac{\sum_{n=0}^{C}n z^{n}}{\sum_{n=0}^{C}z^{n}}.
\label{densityfugacity} 
\end{align} 
Relation (\ref{densityfugacity}) is shown in Fig.\ \ref{fig:densityfugacityI}. Depending on the value of $q/u$, we distinguish three qualitatively different scenarios.

%%%%%%%%%%%%%%%%%%%%%%%%%%%%%%%%%%%%%%%%%%%%%%%%%%%%%%%%%%%%%%%%%%%%%%%%%%%%%%%%%%%%%%%%
\begin{figure}[t!] 
\includegraphics[width=0.48\textwidth]{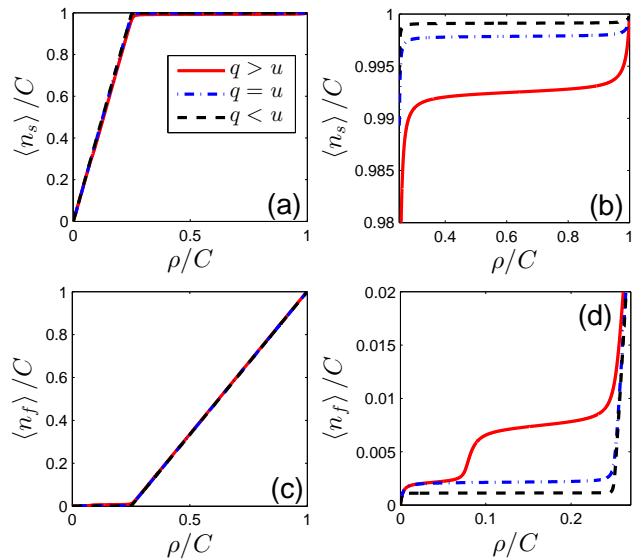}
\centering
\caption{\label{fig:meansequilibrium} (Color online) The scaled mean occupancies of a slow site, $\left<n_{s}\right>/C$, (panels (a), (b)) and the scaled mean occupancies of a fast site $\left<n_{f}\right>/C$ (panels (c), (d)) as the functions of the particle density for $u=0.3$, $C=200$, $M=60$,  $L_{1}/L=1/4$ and for three different values of $q$: $q=u/2$ (the dashed line), $q=u$ (the dot-dashed line), $q=2u$ (the solid line). Panels (b) and (d) show enlargements of regions near phase transitions. In panels (a), (c) all three curves nearly coincide. Averages are calculated analytically using the grand canonical distribution. The fugacity-density relation, $z=z(\rho)$, is obtained by a numerical inversion of (\ref{densityfugacity}).}
\end{figure} 
%%%%%%%%%%%%%%%%%%%%%%%%%%%%%%%%%%%%%%%%%%%%%%%%%%%%%%%%%%%%%%%%%%%%%%%%%%%%%%%%%%%%%%%%

When $q>u$ we observe three continuous transitions between plateaus of $\rho(z)/C$ which sharpen as $C$ is increased. The first transition occurs (approximately) at $z_{\rm c1}=u$.  The height of the transition on the $\rho(z)/C$ axis is proportional to the ratio $(L_{1}M)/(CL)$.  As the fugacity increases through $z_{c1}=u$, the condensate forms on all sites of the slow domain and, at the same time, the mean occupancy of the fast domain saturates (see Fig.\ \ref{fig:meansequilibrium}). The second transition takes place around $z_{c2}=q$ (its height is proportional to $(L_{1}/L)(1-M/C)$). After this transition the mean occupation of the fast domain slightly increases (Fig.\ \ref{fig:meansequilibrium} (d)) while the condensate still growths on slow sites. Further increase of $z$ through $z_{c3}=1$ forces particles to fill up also the fast domain whereas the average occupation of the slow sites saturates. All three transitions  
are of the same type as the Bose-Einstein condensation in the infinite-capacity ZRP. The transitions at $z_{c1}$, $z_{c2}$, correspond ot the Bose-Einstein condensation of particles on the slow sites. Around $z_{c3}=1$, we observe the condensation of \emph{vacancies at fast sites}. Notice that the dynamics of vacancies is dual to that of the particles in the sense that the hopping rate of a vacancy depends on the occupation of the \emph{arrival site}. The model with such hopping rules (dual to ZRP) and with infinite capacities of sites  was studied in \cite{GodrecheTarget}. See also Ref.\ \cite{Blythe} where, similarly to the present case (but in a different model), an extensive number of microscopic condensates was observed. 

\begin{figure}[t!]
\includegraphics[width=0.48\textwidth]{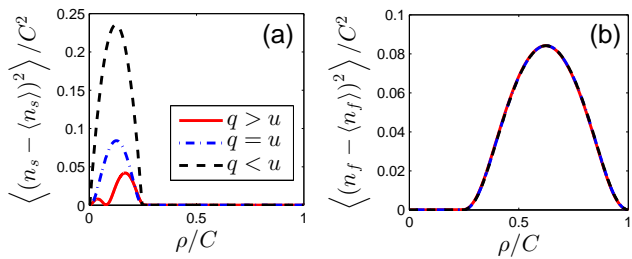}
\centering
\caption{\label{fig:variances} (Color online) Scaled variances of the occupation number of a single slow site (the panel (a)) and of a single fast site (the panel (b)) versus the scaled particle density for $u=0.3$, $C=200$, $M=60$, $L_{1}/L=1/4$, and for three different values of $q$: $q=u/2$ (the dashed line), $q=u$ (the dot-dashed line), $q=2u$ (the solid line). In the panel (b) all three curves coincide. The averages are calculated as described in the caption of Fig.\ \ref{fig:meansequilibrium}.} 
\end{figure}

The second scenario, when $q=u$, is marginal (the dot-dashed line in Figs.\ \ref{fig:densityfugacityI}-\ref{fig:variances}). It can be understood as $q\to u$ limit of the above case. Now, only two continuous transitions occur when the fugacity increases through the values $u$, $1$, respectively. 

When $q<u$, a qualitatively different phase transition occurs at $z_{c2}=q$. The phase transition becomes discontinuous in the limit of both large $C$ and large $M$, $M/C$ fixed. In this limit the transition corresponds to the spontaneous breaking of translation symmetry within the slow domain. This is also illustrated by relatively strong fluctuations of $n_{s}$ depicted in Fig.\ \ref{fig:variances} (a). The number of single-site condensates formed on the slow domain increases as we increase the fugacity  within the interval $z \in (q,1)$. For $z > 1$, the average occupation of each slow site is very close to $C$ (see the dashed line in Fig.\ \ref{fig:meansequilibrium} (b)). On the other hand, when $C$ is large but $M$ is small, $M\sim O(1)$, the transition at $z_{c2}=q$ remains continuous in $C\to\infty$ limit.

%%%%%%%%%%%%%%%%%%%%%%%%%%%%%%%%%%%%%%%%%%%%%%%%%%%%%%%%%%%%%%%%%%%%%%%%%%%%%%%%%%%%%%%%
%%%%%%%%%%%%%%%%%%%%%%%%%%%%%%%%%%%%%%%%%%%%%%%%%%%%%%%%%%%%%%%%%%%%%%%%%%%%%%%%%%%%%%%% 

Strictly speaking, sharp phase transitions occur only in $C\to\infty$ limit. This limit, however, should be understood as follows. The site capacities $C$ can be made arbitrarily large \emph{but not infinite}. Otherwise, i.e., by taking  $C\to\infty$ limit in Eq.\ (\ref{densityfugacity}), we would never reach the phases corresponding to values of $z$ larger than $\min(q,u)$. This stems from the fact that $C\to\infty$ limit of the right-hand side of  Eq.\ (\ref{densityfugacity})  is finite only for $0< z < \min(q,u)$.

%%%%%%%%%%%%%%%%%%%%%%%%%%%%%%%%%%%%%%%%%%%%%%%%%%%%%%%%%%%%%%%%%%%%%%%%%%%%%%%%%%%%%%%%%%%%%%%%%%%%%%%%%%%%%%%%%%%%%
%%%%%%%%%%%%%%%%%%%%%%%%%%%%%%%%%%%%%%%%%%%%%%%%%%%%%%%%%%%%%%%%%%%%%%%%%%%%%%%%%%%%%%%%%%%%%%%%%%%%%%%%%%%%%%%%%%%%%
\subsection{Dynamical self-blocking during condensate growth}
\label{Dynamics}
The hard constraints (\ref{capacity}) on capacities of individual sites lead to a kinetic jamming during the nonequilibrium condensate growth. 
Let us now illustrate this phenomenon for symmetric particle hopping with rates (\ref{BErates}) and for the case $q < u<1$ (for previous studies of the dynamics of ZRP see e.g. Refs. \cite{DynSchutz, GodrecheDynamics, GodrecheLuckDynamics, DynEvans, Juntenen}). 

\begin{figure}[t!]
\includegraphics[width=0.48\textwidth]{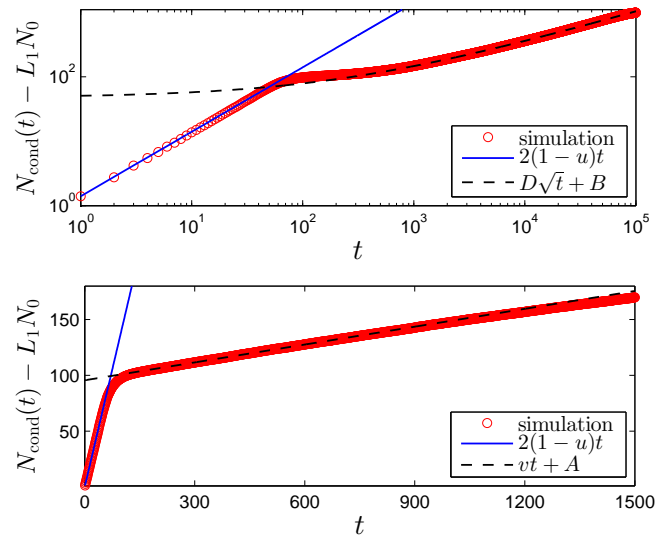}
\caption{\label{fig:relaxation} (Color online)  Growth of the mean condensate size for $C = 100$, $M=99$, $L_{1}=20$, $L=80$, $N_{0}=50$, $u=0.3$, $q=u/10$. The fitting parameters are $D \approx 3.12$, $B  \approx 47.52$, $v \approx 0.05$, $A\approx  95.58$. Numerical data are averaged over $1000$ Monte Carlo runs.}
\end{figure}

At the initial time, $t\!=\!0$, the lattice is half-filled by the particles with $N_{0}\!=\!50$ being the initial number of particles at any site ($C\!=\!100$). We are interested in the evolution of the mean condensate size, $N_{\rm cond}(t)$, defined as the average total number of particles located on the slow sites: 
\begin{equation} 
N_{\rm cond}(t) = \sum_{i=1}^{L_{1}} \left< n_{i}(t) \right>.
\end{equation}
The function $N_{\rm cond}(t)$ obtained from kinetic Monte Carlo simulations is shown in  Fig.~\ref{fig:relaxation}. 

%%%%%%%%%%%%%%%%%%%%%%%%%%%%%%%%%%%%%%%%%%%%%%%%%%%%%%%%%%%%%%%%%%%%%%%%%%%%%%%%%%%%%%%%
%%%%%%%%%%%%%%%%%%%%%%%%%%%%%%%%%%%%%%%%%%%%%%%%%%%%%%%%%%%%%%%%%%%%%%%%%%%%%%%%%%%%%%%%

\begin{figure}[t] 
\centering
\includegraphics[width=.48\textwidth]{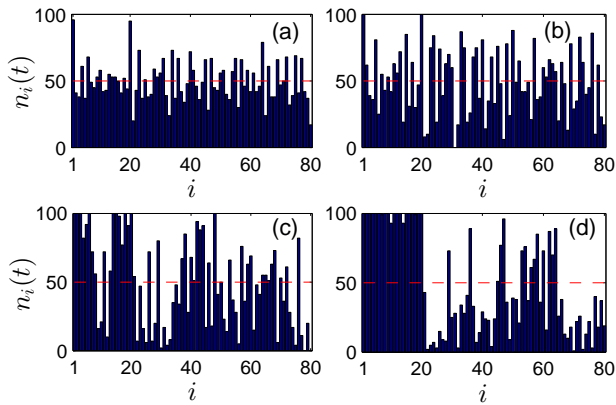}% Here is how to import EPS art
\caption{\label{fig:snapshots} (Color online) Typical system configurations within four dynamical regimes (a) $t=55$, (b) $t=400$, (c) $t=2\times 10^{4}$, (d) $t= 10^{5}$, for $C=100$, $M=99$, $L_{1}=20$, $L=80$, $N_{0}=50$, $u=0.3$, $q=u/10$. The red dashed line shows initial homogeneous distribution of particles.} 
\end{figure}

The condensate growth  starts at the boundary sites and proceeds inwards the slow domain. At small times, primarily the particles initially located at the boundary sites of the fast domain contribute to the condensate growth. On average, $(1\!-\!u)$ particles hops from the site $L$ to the site $1$  per unit time, the same holds true for the sites $L_{1}\!+\! 1$ and $L_{1}$, and hence we observe the linear growth 
\begin{equation} 
\label{smalltimes} 
N_{\rm cond}(t)\approx N_{0}L_{1}+2(1-u)t.
\end{equation}
A typical system configuration within this regime is shown in Fig.\ \ref{fig:snapshots} (a). When the boundary sites of the slow domain are fully occupied, the condensate growth considerably slows down. On intermediate timescales the simulated time-dependence can be fitted by the linear formula   
\begin{equation} 
\label{intermediate} 
N_{\rm cond}(t)\approx N_{0}L_{1} +A+v t, 
\end{equation} 
with $v\! \ll \! 2(1\!-\!u)$ (cf. the lower panel in Fig.\ \ref{fig:relaxation} and the plateau-like part in the upper panel). Within this regime the slow domain is separated from the fast domain by single-site condensates formed at the boundary sites (Fig.\ \ref{fig:snapshots} (b)). As the particles leak through theses blockages, additional ``layers'' of condensate grow on the boundaries of the slow domain (Fig.\ \ref{fig:snapshots} (c)) which eventually yields the slower diffusion-limited growth: 
\begin{equation} 
\label{diffgrowth}
N_{\rm cond}(t)  \approx N_{0}L_{1} + B + D \sqrt{t}. 
\end{equation} 
In this regime, the vacancies diffuse out of the slow domain and the particles join the condensate by the diffusion from the inner sites of the fast domain. After that the condensate size saturates at its equilibrium value and the equilibration of the fast domain follows (Fig.\ \ref{fig:snapshots} (d)). 

When $q\! \geq\! u$, the observed dynamical self-blocking is suppressed, the intermediate regime (\ref{intermediate}) is no longer observed and the relaxation time is much shorter (not shown). On the other hand, if we decrease $M$ (for a given $q$, $q<u$), the self-blocking becomes more pronounced. 

%%%%%%%%%%%%%%%%%%%%%%%%%%%%%%%%%%%%%%%%%%%%%%%%%%%%%%%%%%%%%%%%%%%%%%%%%%%%%%%%%%%%%%%%
%%%%%%%%%%%%%%%%%%%%%%%%%%%%%%%%%%%%%%%%%%%%%%%%%%%%%%%%%%%%%%%%%%%%%%%%%%%%%%%%%%%%%%%%

%%%%%%%%%%%%%%%%%%%%%%%%%%%%%%%%%%%%%%%%%%%%%%%%%%%%%%%%%%%%%%%%%%%%%%%%%%%%%%%%%%%%%%%%%%%%%%%%%%%%%%%%%%%%%%%%%%%%%
%%%%%%%%%%%%%%%%%%%%%%%%%%%%%%%%%%%%%%%%%%%%%%%%%%%%%%%%%%%%%%%%%%%%%%%%%%%%%%%%%%%%%%%%%%%%%%%%%%%%%%%%%%%%%%%%%%%%%
\section{Concluding remarks}
\label{Conclusion}

Let us now summarize the main points in which the present paper goes beyond the previous studies. As for the steady state: A) we have shown that ZRP with finite site capacities has factorized steady state provided the symmetric particle hopping is assumed. For asymmetric dynamics the steady state factorizes if we relax the assumption of the nearest neighbor particle hopping and we allow the particles to overjump jammed sites. B) For finite site capacities, in the limit of large system size the equivalence of ensembles holds \emph{in all phases}. Thus finite site capacities regularize the grand canonical ensemble which, frequently fails to describe the condensed phase in infinite-capacity ZRP. C) On the particular model (hopping rates  (\ref{BErates})) we have demonstrated that the system with \emph{arbitrary large but finite} site capacities possesses a richer phase structure than its counterpart with  \emph{a priori infinite} capacities of sites. As for the dynamics of condensation,  the finite site capacities lead to a dynamical self-blocking during the condensate growth. Detailed analysis of individual dynamical regimes for the model with rates (\ref{BErates}) is given. All these findings suggest several courses of action. In particular, it would be interesting to study physical effects induced by the finite site capacities in realm of more general transport models with factorized steady states \cite{EvansZRP1, Chleboun, WaclawSopik1, WaclawSopik2}.

%%%%%%%%%%%%%%%%%%%%%%%%%%%%%%%%%%%%%%%%%%%%%%%%%%%%%%%%%%%%%%%%%%%%%%%%%%%%%%%%%%%%%%%%%%%%%%%%%%%%%%%%%%%%%%%%%%%%%%%%%%%%%%%%%%%%%%%%%%%%%%%%%%%%%%%%%%%%%%%%%%%%%%%%%%%%%%%%
\section*{Acknowledgments}
Support of this work by the Charles University Grant Agency (grant No.\ 301311), and by the Charles University in Prague (project No.\ SVV-2014-267-305) is gratefully acknowledged. 
%%%%%%%%%%%%%%%%%%%%%%%%%%%%%%%%%%%%%%%%%%%%%%%%%%%%%%%%%%%%%%%%%%%%%%%%%%%%%%%%%%%%%%%%%%%%%%%%%%%%%%%%%%%%%%%%%%%%%%%%%%%%%%%%%%%%%%%%%%%%%%%%%%%%%%%%%%%%%%%%%%%%%%%%%%%%%%%%
\bibliography{references}
%%%%%%%%%%%%%%%%%%%%%%%%%%%%%%%%%%%%%%%%%%%%%%%%%%%%%%%%%%%%%%%%%%%%%%%%%%%%%%%%%%%%%%%%%%%%%%%%%%%%%%%%%%%%%%%%%%%%%%%%%%%%%%%%%%%%%%%%%%%%%%%%%%%%%%%%%%%%%%%%%%%%%%%%%%%%%%%%
\end{document}